## ARTICLE

# Pseudogap phase of cuprate superconductors confined by Fermi surface topology

N. Doiron-Leyraud[1], O. Cyr-Choinière[1], S. Badoux[1], A. Ataei[1], C. Collignon[1], A. Gourgout[1], S. Dufour-Beauséjour [1], F.F. Tafti[1], F. Laliberté[1], M.-E. Boulanger[1], M. Matusiak[1,2], D. Graf[3], M. Kim[4,5], J.-S. Zhou[6], N. Momono[7], T. Kurosawa[8], H. Takagi[9] & Louis Taillefer[1,10]

The properties of cuprate high-temperature superconductors are largely shaped by competing phases whose nature is often a mystery. Chiefly among them is the pseudogap phase, which sets in at a doping $p^*$ that is material-dependent. What determines $p^*$ is currently an open question. Here we show that the pseudogap cannot open on an electron-like Fermi surface, and can only exist below the doping $p_{FS}$ at which the large Fermi surface goes from hole-like to electron-like, so that $p^* \leq p_{FS}$. We derive this result from high-magnetic-field transport measurements in $La_{1.6-x}Nd_{0.4}Sr_xCuO_4$ under pressure, which reveal a large and unexpected shift of $p^*$ with pressure, driven by a corresponding shift in $p_{FS}$. This necessary condition for pseudogap formation, imposed by details of the Fermi surface, is a strong constraint for theories of the pseudogap phase. Our finding that $p^*$ can be tuned with a modest pressure opens a new route for experimental studies of the pseudogap.

[1] Institut Quantique, Département de Physique & RQMP, Université de Sherbrooke, Sherbrooke, QC J1K 2R1, Canada. [2] Institute of Low Temperature and Structure Research, Polish Academy of Sciences, 50-422 Wrocław, Poland. [3] National High Magnetic Field Laboratory, Florida State University, Tallahassee, FL 32306, USA. [4] École Polytechnique, CNRS, Université Paris-Saclay, 91128 Palaiseau, France. [5] Collège de France, 75005 Paris, France. [6] Materials Science and Engineering Program/Mechanical Engineering, University of Texas - Austin, Austin, TX 78712, USA. [7] Department of Applied Sciences, Muroran Institute of Technology, Muroran 050-8585, Japan. [8] Department of Physics, Hokkaido University, Sapporo 060-0810, Japan. [9] Department of Advanced Materials, University of Tokyo, Kashiwa 277-8561, Japan. [10] Canadian Institute for Advanced Research, Toronto, ON M5G 1Z8, Canada. Correspondence and requests for materials should be addressed to N.D.-L. (email: Nicolas.Doiron-Leyraud@USherbrooke.ca) or to L.T. (email: Louis.Taillefer@USherbrooke.ca)

A central puzzle of cuprate superconductors[1], the pseudogap is a partial gap that opens in their spectral function, detected most directly by angle-resolved photoemission spectroscopy (ARPES). It opens below a temperature $T^*$ that decreases monotonically with increasing hole concentration (doping) $p$. For example, $T^* = 130 \pm 20$ K in La$_{2-x}$Sr$_x$CuO$_4$ (LSCO) at $p = 0.15$[2] and $T^* = 75 \pm 5$ K in La$_{1.6-x}$Nd$_{0.4}$Sr$_x$CuO$_4$ (Nd-LSCO) at $p = 0.20$[3]. Transport properties like the electrical resistivity $\rho(T)$ and the Nernst coefficient $\nu(T)$ are affected by the opening of the pseudogap and so may be used to detect $T^*$, as previously reported in refs. [4,5] and ref. [6], respectively.

In Fig. 1a, we show the temperature-doping phase diagram of LSCO, Nd-LSCO and La$_{1.8-x}$Eu$_{0.2}$Sr$_x$CuO$_4$ (Eu-LSCO). We see that all three materials have the same $T^*$ up to $p \sim 0.17$, irrespective of their different crystal structures[6]. Indeed, Nernst data[6] show, for example, that $T^* = 120 \pm 10$ K at $p = 0.15$ in both LSCO and Nd-LSCO, whose structure in that part of the phase diagram is orthorhombic, and $T^* = 115 \pm 10$ K at $p = 0.16$ in Eu-LSCO, whose structure in that part of the phase diagram is tetragonal (LTT) (Supplementary Fig. 1). In Nd-LSCO and Eu-LSCO, $T^*$ decreases linearly all the way from $p \sim 0.08$ to $p \sim 0.23$ (blue line in Fig. 1a). On its trajectory, the $T^*$ line goes unperturbed through the LTT transition at $p \sim 0.14$ for Eu-LSCO (short green line).

at $p \sim 0.19$ for Nd-LSCO (short red line). Clearly, the pseudogap mechanism does not care about the crystal structure (Note that it is also robust against disorder[7]).

The critical doping $p^*$ at which the pseudogap phase comes to an end, however, is material-specific. In LSCO, the linear decrease in $T^*$ vs $p$ comes to an end at $p^* = 0.18 \pm 0.01$[8,9]. In Nd-LSCO, resistivity measurements[4,5] at $p = 0.20$ and above show that the $T^*$ line only comes to an end at $p^* = 0.23 \pm 0.01$. Why does $T^*$ not continue to track the dashed blue line in Fig. 1 beyond $p = 0.17$ for LSCO, or beyond $p = 0.23$ for Nd-LSCO? In Fig. 1e, we plot $p^*$ for the three single-layer cuprates LSCO, Nd-LSCO and Bi2201[10], as a function of $p_{FS}$, the doping at which the Fermi surface undergoes a change from hole-like to electron-like as determined by ARPES measurements[2,3,11,12]. Within error bars, we observe that $p^* = p_{FS}$, in other words, it appears that what limits $p^*$ is the constraint that the pseudogap cannot open on an electron-like Fermi surface.

In the following, we examine whether this connection is accidental or not, by independently probing how $p^*$ and $p_{FS}$ evolve under the effect of hydrostatic pressure in Nd-LSCO. With a maximal $T_c$ of only 20 K, this cuprate provides a window into the pseudogap phase near its end point at $p^*$, free of superconductivity, down to the $T = 0$ limit, achieved by applying a magnetic field of 30 T (or greater), whose sole effect is to suppress superconductivity and reveal the underlying normal state (ref. [5]). Our measurements reveal a large and unexpected downward shift of $p^*$ with pressure, which we find to be driven by a corresponding change in $p_{FS}$, so that the relation $p^* \le p_{FS}$ is obeyed. This fundamental property has direct and fundamental implications for the mechanism of pseudogap formation.

## Results

**Determination of $p^*$ and $p_{FS}$.** Our study is based on transport signatures of $p^*$ and $p_{FS}$. We begin with $p^*$. When $p < p^*$, the electrical resistivity $\rho(T)$, Nernst coefficient and Hall coefficient $R_H(T)$ all exhibit large upturns at low temperature[4,5,6,9]—

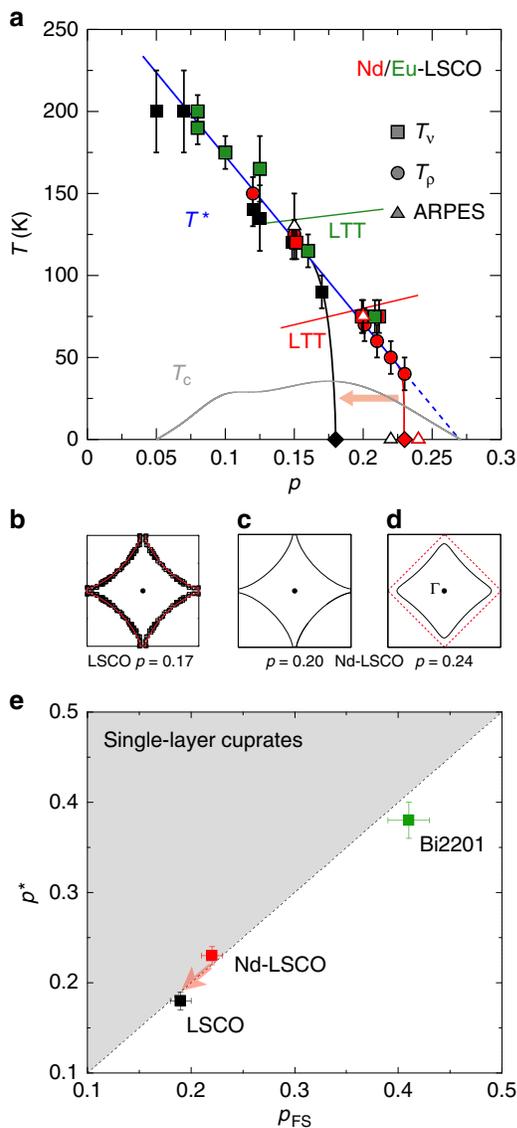

**Fig. 1** Phase diagram and correlation between $p^*$ and $p_{FS}$. **a** Temperature-doping phase diagram, showing the pseudogap temperature $T^*$ measured by ARPES (triangles), electrical resistivity (circles) and Nernst effect (squares), in LSCO (black), Nd-LSCO (red) and Eu-LSCO (green) (adapted from ref. [6]; see references therein). The diamonds mark the critical doping $p^*$ at which the pseudogap phase ends at $T = 0$ (in the absence of superconductivity), for LSCO (black; $p^* = 0.18$, refs. [8,9]) and Nd-LSCO (red; $p^* = 0.23$, refs. [4,5]). The short solid lines mark the transition into the low-temperature tetragonal structure (LTT) for Eu-LSCO (green; ref. [36] and Nd-LSCO (red; ref. [34]), in the interval where it crosses $T^*$. The grey line is the superconducting critical temperature $T_c$ of LSCO. The dashed blue line is a linear extension of the $T^*$ line (blue). The red arrow illustrates the effect of applying hydrostatic pressure to Nd-LSCO: it shifts $p^*$ down in doping. **b, c, d** Sketch of the Fermi surface in the first Brillouin zone (frame), as measured by ARPES in LSCO (ref. [11]) at $p = 0.17$ (**b**) and in Nd-LSCO (ref. [3]) at $p = 0.20$ (**c**) and $p = 0.24$ (**d**). The red dashed line is the anti ferromagnetic zone boundary (AFZB), which intersects the hole-like Fermi surface of Nd-LSCO at $p = 0.20$, but not its electron-like surface at $p = 0.24$. **e** Correlation between $p^*$ and $p_{FS}$ in single-layer cuprates. $p^*$ is measured in the normal state at $T = 0$, by high-field transport in Nd-LSCO ($p^* = 0.23 \pm 0.01$, ref. [5]) and LSCO ($p^* = 0.18 \pm 0.01$, refs. [8,9]), and by high-field NMR in Bi2201 ($p^* = 0.38 \pm 0.02$, ref. [10]). $p_{FS}$ is measured by ARPES in LSCO ($p_{FS} = 0.19 \pm 0.01$, refs. [2,11]), Nd-LSCO ($p_{FS} = 0.22 \pm 0.01$; ref. [3]) and Bi2201 ($p_{FS} = 0.41 \pm 0.02$, ref. [12]). The red arrow shows the effect of applying pressure to Nd-LSCO (red square). We find that $p^*$ and $p_{FS}$ decrease in tandem, preserving the equality $p^* = p_{FS}$ and thereby showing that $p^*$ is constrained by the condition $p^* \le p_{FS}$. The grey shading marks the forbidden region

signatures of the pseudogap phase, attributed to a drop in carrier density $n$ from $n = 1 + p$ above $p^\star$ to $n = p$ below $p^\star$[5,9,13]. In Nd-LSCO at $p = 0.20$, ARPES sees a gap opening at $T^\star = 75$ K (ref. 3), precisely the temperature below which the resistivity exhibits an upward deviation from its high-$T$ linear behaviour[4,5]. By contrast, at $p = 0.24$ the three transport coefficients show no trace of any upturn at low $T$[4,5,6], with the resistivity remaining linear down to $T \rightarrow 0$ (Fig. 2a), consistent with the absence of a gap in ARPES data[3].

In Fig. 3, we reproduce published data[5] for $\rho(T)$ (Fig. 3a) and $R_H(T)$ (Fig. 3c) in Nd-LSCO at different dopings. The upturns decrease as $p$ approaches $p^\star$ from below. We determine $p^\star$ as the doping where the upturns in $\rho(T)$ and $R_H(T)$ vanish, giving $p^\star = 0.23 \pm 0.01$ (Fig. 4). (Note that at that doping $\rho(T)$ displays a slight upturn while $R_H(T)$ remains flat. This difference comes possibly from the fact that $\rho(T)$ is sensitive to the total carrier density while $R_H(T)$ is balanced by electron- and hole-like contributions, as in the reconstructed Fermi surface just below $p^\star$ for an anti ferromagnetic scenario[14]. See discussion in ref. 5.)

We now turn to $p_{FS}$. In a single-layer cuprate, the Fermi surface changes topology from hole-like at $p < p_{FS}$ (Fig. 1c) to electron-like at $p > p_{FS}$ (Fig. 1d). ($p_{FS}$ is the doping at which the van Hove singularity in the density of states crosses the Fermi level.) ARPES studies show that the Fermi surface is hole-like in LSCO at $p = 0.17$[11] and Nd-LSCO at $p = 0.20$[3], while it is electron-like at $p = 0.20$[2] and $p = 0.24$[3], respectively, so that $p_{FS} = 0.19 \pm 0.01$ in LSCO and $p_{FS} = 0.22 \pm 0.01$ in Nd-LSCO.

**Changes under hydrostatic pressure.** We now examine the effect of pressure on $p_{FS}$ and $p^\star$. Pressure is known to change the crystal structure of Nd-LSCO from LTT at ambient pressure to HTT at $P > 4.2$ GPa[15]. Our band–structure calculations based on a standard tight-binding model (see 'Methods' section and Supplementary Fig. 2) show that this causes a decrease in the ratio $|t'/t|$, where $t$ and $t'$ are nearest- and next-nearest-neighbour hopping parameters. It is a property of this model that the Fermi surface goes from hole-like to electron-like with decreasing $|t'/t|$ (at fixed $p$), consistent with fits to ARPES data on LSCO[2,11]. Pressure applied to Nd-LSCO is therefore expected to reduce $p_{FS}$ (Fig. 5c). Although the maximum pressure in our experiment is 2.0 GPa, it still reduces the $CuO_6$ octahedron tilt angle of the LTT structure significantly towards the HTT phase[15], which decreases $|t'/t|$.

Experimentally, we study the pressure dependence of $p_{FS}$ by looking at $R_H$ for $p \geq p^\star$. In LSCO in the regime above $p_{FS} \sim 0.19$, $R_H$ decreases linearly with doping as the system moves away from $p_{FS}$ (Supplementary Fig. 3), to eventually become negative above $p \sim 0.35$[16,17]. Quantitatively, an increase in doping by $\delta p = 0.02$ corresponds to a ~13% drop in $R_H$ (Fig. 5b). Since all there is in this regime is a single large electron-like Fermi surface, Nd-LSCO in the regime above $p_{FS} \sim 0.22$ must display a similar linear

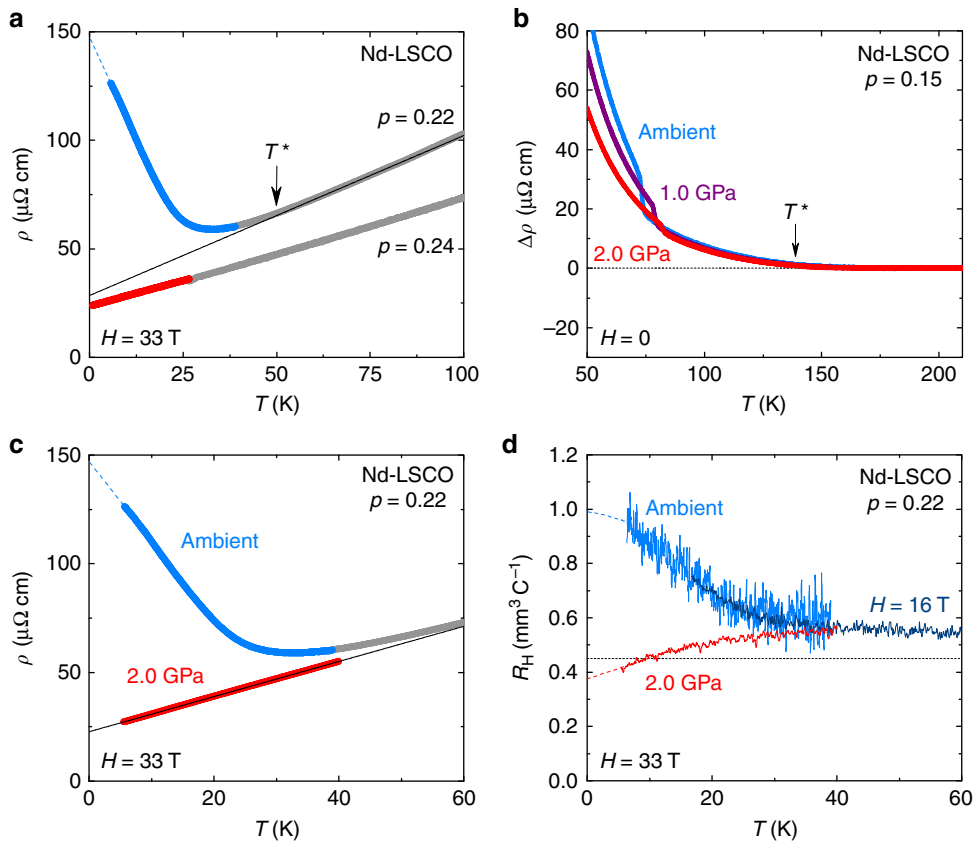

**Fig. 2** Effect of pressure on Nd-LSCO. Transport properties of Nd-LSCO as a function of temperature, at different dopings as indicated. In **a**, **c** and **d**, grey data are in zero field and coloured data are in magnetic field. **a** Resistivity of Nd-LSCO at $p = 0.22$ from ref. 5 (blue) and at $p = 0.24$ from ref. 4 (red), in $H = 33$ T and ambient pressure. At $p = 0.22$, $\rho(T)$ is seen to deviate upwards from its linear $T$ dependence at high $T$ (black line) below the onset temperature $T^\star$ of the pseudogap phase (Fig. 1a). **b** Difference between the data and the linear fits of Supplementary Fig. 6, showing that the onset of the upturn at $T^\star$ (arrow) is independent of pressure. The sharp kink is due to the structural transition from orthorhombic at high $T$ (LTO) to tetragonal at low $T$ (LTT) (see Fig. 1). **c** Resistivity of Nd-LSCO at $p = 0.22$ and $H = 33$ T, at ambient pressure (blue, ref. 5) and for $P = 2.0$ GPa (red, this work), with a linear fit (black line) to the $P = 2.0$ GPa data. **d** As in **c**, but for the Hall coefficient $R_H(T)$. The horizontal dotted line is the value expected from the large hole-like Fermi surface. Coloured dashed lines are a guide to the eye

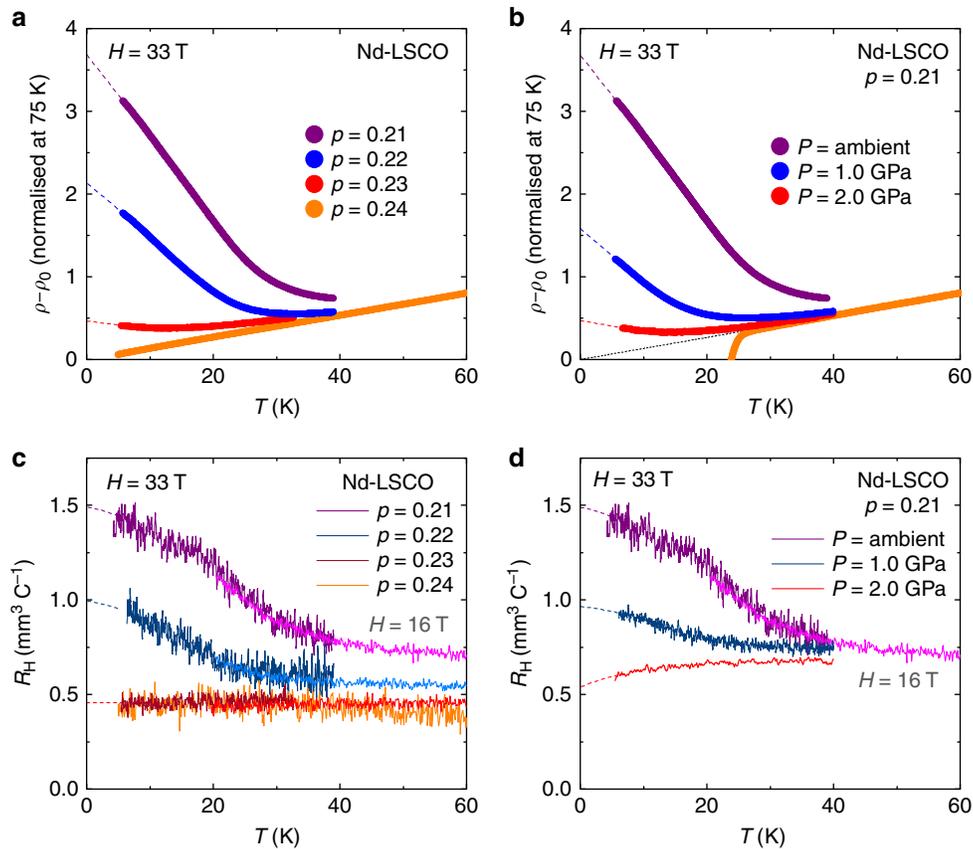

**Fig. 3** Effect of doping and pressure on Nd-LSCO. Comparing the effect of doping on the pseudogap phase in Nd-LSCO (left panels) to the effect of pressure (right panels). **a, b** Electrical resistivity expressed as $\Delta\rho = \rho(T) - \rho_0$ (normalised at 75 K), where $\rho_0$ is the residual resistivity estimated from a linear fit to $\rho(T)$ above $T^*$ (ref. [5]). The values of $\Delta\rho$ at $T \to 0$ are plotted vs $p$ in Fig. 4a. **a** At ambient pressure, for various dopings as indicated, in $H = 33$ T (data from ref. [5]). **b** At $p = 0.21$, for different pressures as indicated, in $H = 33$ T. Data at $P = 2.0$ GPa and $H = 0$ are also shown (orange), together with a linear extrapolation to $T = 0$ (black dotted line). **c, d** Same as in **a** and **b** but for the Hall coefficient $R_H$. Light and dark coloured curves are in $H = 16$ and 33 T, respectively. The values at $T \to 0$, labelled $R_H(0)$, are plotted vs $p$ in Fig. 4b. All dashed lines are a guide to the eye

decrease of $R_H$ with doping, possibly with a different absolute value. Hence, the same relative change in $R_H$ is expected to correspond to a similar displacement in doping. In Fig. 5a (see also Supplementary Fig. 4), we see that $R_H$ in Nd-LSCO at $p = 0.25$ decreases by $\sim 12$% under 2 GPa, which implies that $p_{FS}$ in Nd-LSCO shifts down by $\delta p_{FS} \approx 0.02$ under 2 GPa, based on the LSCO data.

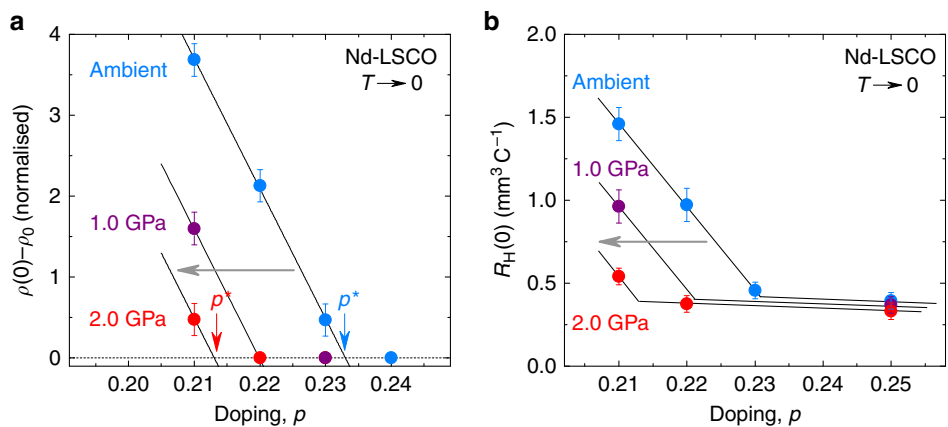

**Fig. 4** Effect of pressure on $p^*$. **a** Magnitude of the upturn in $\rho(T)$ at $T \to 0$ and for $H = 33$ T as a function of doping, for ambient pressure (blue), $P = 1.0$ (purple) and 2.0 GPa (red), obtained from ambient pressure data (Fig. 3a) and from pressure data at $p = 0.21$ (Fig. 3b). The pseudogap critical point $p^*$ is where the upturn goes to zero (arrow). **b** $R_H(0)$ vs doping for ambient pressure (blue), $P = 1.0$ (purple) and 2.0 GPa (red), obtained from ambient pressure data at $p = 0.21$, 0.22 and 0.23 (Fig. 3c) for $H = 33$ T, and $p = 0.25$ (Supplementary Fig. 4) for $H = 16$ T, and from pressure data at $p = 0.21$ (Fig. 3d) and $p = 0.22$ (Fig. 2d) for $H = 33$ T, and $p = 0.25$ (Supplementary Fig. 4) at $H = 16$ T. All lines are a guide to the eye. The effect of pressure is to shift $p^*$ down in doping (grey arrows), by roughly $dp^* = -0.02$ for 2.0 GPa. All the error bars reflect the uncertainty on the extrapolation to $T = 0$

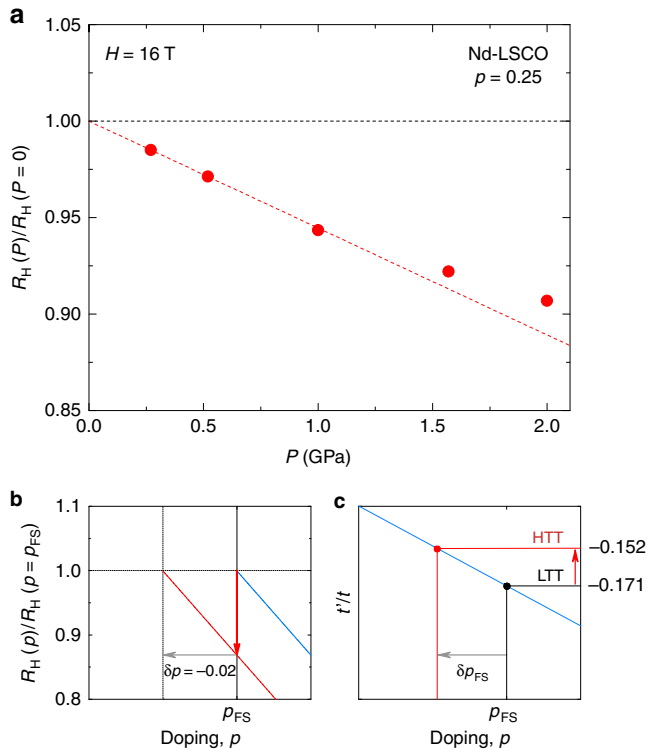

**Fig. 5** Effect of pressure on $p_{FS}$. **a** Pressure dependence of the Hall coefficient in Nd-LSCO at $p = 0.25 > p^*$, normalised by its value at ambient pressure, $R_H(0) = 0.4$ mm$^3$ C$^{-1}$ (Supplementary Fig. 4). The dashed line is a linear fit through the data at low pressure, whose slope is $-6$ % GPa$^{-1}$. **b** Sketch of the doping dependence of $R_H$ in LSCO (Supplementary Fig. 3), normalised by its value at $p_{FS}$ (blue line). The red line illustrates the effect of shifting $p_{FS}$ down in doping by an amount $\delta p_{FS} = -0.02$ (grey arrow): $R_H$ decreases by -13 % (red arrow). This is the amount by which $R_H$ decreases in Nd-LSCO under 2.0 GPa, for $p > p^*$ (**a**). **c** Sketch of the theoretical dependence of $p_{FS}$ on the band-structure parameter $t'/t$. For a given value of $t'/t$, $p_{FS}$ is given by the blue line, which separates the regions where the cuprate Fermi surface is electron-like (above) and hole-like (below). Our band-structure calculations for Nd-LSCO yield $t'/t = -0.171$ in the LTT structure (black) and $t'/t = -0.152$ in the HTT structure (red) (Supplementary Fig. 2), so that $p_{FS}$ is expected to decrease (grey arrow) under pressure, as observed experimentally

To study the pressure dependence of $p^*$, we measured the high-field normal-state resistivity under pressure up to 2 GPa. Our data on LSCO below $p^*$ show that the pseudogap is unaffected by pressure, in the sense that the amplitude of the low-$T$ upturn in the resistivity at $p = 0.143$ (ref. [9]) is unchanged (Supplementary Fig. 5). In Nd-LSCO at $p = 0.15$, our measurements (Fig. 2b and Supplementary Fig. 6) also show that $T^*$ is independent of pressure, as previously observed in YBa$_2$Cu$_3$O$_y$ (ref. [18]). So a pressure of 2 GPa does not tune $T^*$ directly. Nevertheless, in Nd-LSCO just below $p^*$ we observe a dramatic effect of pressure on the pseudogap: at $p = 0.22$, 2.0 GPa completely eliminates the low-$T$ upturn in the resistivity (Fig. 2c), resulting in a linear $T$ behaviour characteristic of the regime at $p^*$ and above (refs. [4,8]). A suppression of the low-$T$ upturn is also seen at $p = 0.21$ (Fig. 3b). This large and unexpected effect of pressure on the pseudogap is our main experimental finding.

In Fig. 3, we compare the effect of pressure on the resistivity and Hall coefficient of Nd-LSCO with the effect of doping. (Note that unlike for YBa$_2$Cu$_3$O$_y$, pressure does not change the doping in Nd-LSCO, which is set by the Sr content.) We see that pressure has the same effect as doping, consistent with a lowering of $p^*$ induced by pressure. Quantitatively, $p^*$ moves down from 0.23 at $P = 0$ to 0.21 at $P = 2.0$ GPa (Fig. 4). A downward shift $\delta p^* = -0.02$ in 2.0 GPa is the same shift ($\delta p_{FS} \sim -0.02$) that is observed for $p_{FS}$. We infer that it is the downward movement of $p_{FS}$ that constrains $p^*$ to move down with pressure, thereby showing that the condition $p^* \lesssim p_{FS}$ must hold.

## Discussion

This elucidates why the pseudogap phase of LSCO ends at a lower doping than in Nd-LSCO, for indeed $p^* = p_{FS}$ in LSCO, within error bars (Fig. 1e). In most cuprates, $p_{FS}$ is much higher, as in the single-layer material Bi2201, for example, where $p_{FS} = 0.41 \pm 0.02$ (ref. [12]). Remarkably, $p^*$ is nearly as high, with $p^* = 0.38 \pm 0.02$ in Bi2201 (ref. [10]), as illustrated in Fig. 1e.

Our finding that $p^* \lesssim p_{FS}$ is consistent with the known properties of all cuprates. In particular, it holds true for all known single-layer cuprates, including not only those in Fig. 1e, but also Tl$_2$Ba$_2$CuO$_{6+\delta}$, for example, where at $p \sim 0.3$ the Fermi surface is hole-like and there is no pseudogap[19]. It also holds for bi-layer cuprates, such as Bi$_2$Sr$_2$CaCu$_2$O$_{8+\delta}$, in the sense that the pseudogap opens since only once[20] or after[21] both Fermi surfaces (bonding and anti-bonding) have become hole-like[22]. The close proximity of $p^*$ to a van Hove singularity (at $p_{FS}$) may have some impact on the physics of the pseudogap. However, since the velocity vanishes at the van Hove point, we do not expect any singularity in the transport properties. Moreover, in the presence of substantial $c$-axis dispersion, of the order of $t_z \sim 20$ meV in Nd-LSCO, the divergence in the density of states gets cut off at low temperature[23].

It is striking that a minute change in the Fermi surface, smaller than that illustrated between Fig. 1c ($p$ slightly below $p_{FS}$) and Fig. 1d ($p$ slightly above $p_{FS}$), can switch off a gap of magnitude $\sim 20$ meV (ref. [3]). This extreme sensitivity of the pseudogap on the details of the Fermi surface suggests that for the pseudogap to form it is necessary that the Fermi surface intersects the anti ferromagnetic zone boundary (AFZB), the dashed line in Fig. 1d, as proposed in refs. [20,24]. This is precisely what happens when the doping drops below $p_{FS}$. There is empirical evidence that this intersection may indeed be a crucial element of the pseudogap mechanism. First, this AFZB is where in $k$-space the separation of ungapped and (pseudo)gapped states occurs, as detected by quasiparticle interference in scanning tunnelling microscopy[21]. In other words, the AFZB is the pseudogap phase's organising principle in $k$-space: it defines the so-called 'Fermi arcs'. Note that when the pseudogap turns on (with decreasing $T$ or $p$), the $k$-space area contained by the ungapped states (between the Fermi arcs and the AFZB) goes from $A \sim 1 + p$ to $A \sim p$ (ref. [21]). Second, upon crossing $p^*$ from above, the carrier density measured by the Hall number goes from $n = 1 + p$ to $n = p$ (refs. [5,13]), consistent with $A \sim n$. The simplest way to obtain a loss of 1.0 hole per planar Cu atom is to reconstruct the large hole-like Fermi surface (with $1 + p$ holes) by folding it about the AFZB, which produces four small nodal hole pockets (with $p$ holes)[14]. This can be achieved either by AF order with a wavevector $\mathbf{Q} = (\pi, \pi)$ or by an Umklapp surface coincident with the AFZB, as in the YRZ model[25]. If this is indeed how the pseudogap phase transforms the Fermi surface, then no Fermi arcs (or nodal pockets) can form when $p > p_{FS}$. Electron-doped cuprates provide a clear example of Fermi surface reconstruction caused by long-range AF order, where the AFZB plays a key role, but where no pseudogap phase forms[26]. The issue of why there is a pseudogap in hole-doped cuprates and not in electron-doped cuprates, however, is an open question.

If the pseudogap phase needs states near $(\pi, 0)$ to form, the question is at what energy do those states have to be relative to $\varepsilon_F$? We presume that they should be within the pseudogap energy $\Delta_{PG}$ of $\varepsilon_F$. This implies that for $p$ just above $p_{FS}$, the pseudogap phase can still form. This introduces some 'width' to the criterion $p^* \leq p_{FS}$. In a similar vein, a 3D dispersion of the Fermi surface in the $k_z$ (or $c$) direction will also give some 'width' to the criterion, since at a given doping the Fermi surface can be electron-like at some $k_z$ value but still hole-like at some other $k_z$ value.

The requirement that a pseudogap cannot form in a cuprate with an electron-like Fermi surface imposes a stringent constraint on theories of the pseudogap phase[25,27–31]. In the YRZ model[25], a pseudogap forms because carriers undergo Umklapp scattering, inherited from the Mott insulator, at points where the Fermi surface intersects the AFZB; this model therefore agrees with our proposed constraint. Our findings are also broadly consistent with spin-fermion models that hinge on the hot spots that lie at the intersection of Fermi surface and AFZB[27,28]. More specifically, two recent theoretical studies[29,32] find that a pseudogap only opens on hole-like Fermi surfaces and that $p^* \leq p_{FS}$ for a wide range of band–structure parameters, even in the strong-coupling regime where the anti ferromagnetic correlations responsible for the pseudogap are short ranged. On the other hand, in scenarios characterised by a wavevector $\mathbf{Q} = (0, 0)$, the AFZB plays no special role and there is then no obvious reason for the constraint to be effective. This would seem to rule out nematic order[31] and intra-unit cell magnetic order[30,33] as possible drivers of the pseudogap phase. Instead, these orders would be secondary instabilities[6].

On the experimental side, the ability to continuously suppress and restore the pseudogap with pressure in a given sample without changing the doping or disorder level provides a promising avenue to study the pseudogap state, compatible with a range of probes such as Raman, optics, X-ray and neutron scattering.

## Methods

**LSCO samples.** Single crystals of La$_{2-x}$Sr$_x$CuO$_4$ (LSCO) were grown by the flux-zone technique, with nominal Sr concentrations of $x = 0.145$ at Hokkaido University and $x = 0.18$ at the University of Tokyo. Samples for resistivity measurements were cut in the shape of small rectangular platelets, of typical dimensions 1 mm × 2 mm × 0.5 mm, with the smallest dimension along the $c$ axis. Contacts were made using H20E silver epoxy, diffused by annealing. The hole concentration (doping) $p$ for our nominal $x = 0.145$ sample was determined using the doping dependence of the (tetragonal to orthorhombic) structural transition temperature, $T_{LTO}$, which is detected in the resistivity as a small but sharp kink. This yields $p = 0.143$ (ref. [9]). For our sample with $x = 0.18$, we take $p = x$.

**Nd-LSCO samples.** Single crystals of La$_{2-x-y}$Nd$_y$Sr$_x$CuO$_4$ (Nd-LSCO) were grown at the University of Texas at Austin with a Nd content $y = 0.4$, using a travelling-float-zone technique, and cut from boules with nominal Sr concentrations $x = 0.15$, 0.21, 0.22, 0.23 and 0.25. The samples were prepared for transport measurements as described above for the LSCO samples. For all five samples, the hole concentration $p$ is given by $p = x$, with an error bar $\pm 0.003$. The samples labelled here $p = 0.21$, 0.22, 0.23 and 0.24 are the same samples as those studied in ref. [5]; the sample labelled here $p = 0.25$ is a new sample, with $p = 0.25 \pm 0.003$.

**Resistivity and Hall measurements.** The electrical transport measurements were performed via a standard four-point low AC technique using an SR830 lock-in amplifier and a Keithley 6221 current source. A current of typically 2 mA was applied within the CuO$_2$ planes and the magnetic field along the $c$-axis. The longitudinal and Hall resistances $R_{xx}$ and $R_{xy}$ were measured at Sherbrooke in steady fields up to 16 T and at the NHMFL in steady fields up to 45 T. The Hall resistance $R_{xy}$ is obtained by reversing the field and anti-symmetrizing the data, as $R_{xy}(H) = (R_{xy}(+H) - R_{xy}(-H))/2$.

**Application of pressure.** Pressure was applied on our samples using a miniature non-magnetic piston-cylinder cell. The pressure medium is Daphne oil 7474, which remains liquid at all pressures measured here at 300 K, ensuring a high degree of hydrostaticity. The internal pressure is measured both at room temperature and at 4.2 K, using either the fluorescence of a small ruby chip or a Sn manometer. The values quoted throughout are the low temperature pressures. The error bar on all the pressure values is $\pm 0.05$ GPa, which comes from the uncertainty in measuring the position of the fluorescence peaks. For each measurement, the cell was cooled slowly (<1 K min$^{-1}$) to ensure a homogeneous freezing of the pressure medium.

**Band-structure calculations.** Band–structure calculations were performed by using the full potential augmented plane wave band method, implemented in the WIEN2k package. We used the local density approximation for the exchange-correlation potential. We used 1000 k-points inside the first Brillouin zone. The convergence of total energy with respect to the number of k-points was checked to have a precision better than 0.007 eV per formula unit. We calculated the electronic structure of La$_2$CuO$_4$ using the structural parameters of Nd-LSCO measured experimentally by X-ray diffraction for both ambient pressure (LTT structure; ref. [34] and $P = 4.2$ GPa (HTT structure; ref. [15]). In the case of $P = 4.2$ GPa (HTT), in order to determine the internal position of atoms inside the CuO$_6$ octahedron, we assumed an isotropic contraction of the CuO$_6$ octahedra from hydrostatic pressure. Experiments on La$_2$CuO$_4$ have shown this assumption to be valid[35]. Tight-binding hopping parameters were obtained by fitting Cu($d_{x^2-y^2}$)-driven bands for high-symmetry points of the anti ferromagnetic zone boundary.

**Sample size.** No statistical methods were used to predetermine sample size.

**Data availability.** All relevant data are available from the authors.

## Acknowledgements

We thank the following for enlightening discussions: S. Benhabib, G. Boebinger, C. Bourbonnais, J. Chang, S. Chakravarty, M. Ferrero, A. Georges, G. Kotliar, G. G. Lonzarich, A. J. Millis, C. Pépin, C. Proust, A. Sacuto, D. Sénéchal, A.-M. Tremblay, S. Verret, and W. Wu. A portion of this work was performed at the National High Magnetic Field Laboratory, which is supported by National Science Foundation Cooperative Agreement No. DMR-1157490 and the State of Florida. L.T. thanks the Collège de France for its hospitality and the European Research Council (Grant No. ERC-319286 QMAC) and LABEX PALM (ANR-10-LABX-0039-PALM) for their support, while this article was written. J.-S.Z. was supported by DOD-ARMY (W911NF-16-1-0559) in the United States. H.T. acknowledges MEXT Japan for a Grant-in-Aid for Scientific Research. L.T. acknowledges support from the Canadian Institute for Advanced Research (CIFAR) and funding from the Natural Sciences and Engineering Research Council of Canada (NSERC; PIN:123817), the Fonds de recherche du Québec − Nature et Technologies (FRQNT), the Canada Foundation for Innovation (CFI) and a Canada Research Chair. Part of this work was funded by the Gordon and Betty Moore Foundation's EPiQS Initiative (Grant GBMF5306 to L.T.).

## Author contributions

N.D.-L., O.C.-C., S.B., A.A., A.G., S.D.-B., F.F.T., and F.L. performed the transport measurements under pressure and ambient conditions at Sherbrooke. N.D.-L., O.C.-C., S.B., C.C., A.G., S.D.-B., M.-E.B., M.M., and D.G. performed the transport measurements under pressure and ambient conditions at the NHMFL in Tallahassee. M.K. performed the band−structure calculations. J.-S.Z. grew the Nd-LSCO single crystals. N.M. and T.K. grew the LSCO single crystal with $x = 0.145$. H.T. grew the LSCO single crystal with $x = 0.18$. N.D.-L. and L.T. wrote the manuscript, in consultation with all authors. N.D.-L. and L.T. supervised the project.

## Additional information

Supplementary Information accompanies this paper at https://doi.org/10.1038/s41467-017-02122-x.

Competing interests: The authors declare no competing financial interests.

Reprints and permission information is available online at http://npg.nature.com/reprintsandpermissions/

Publisher's note: Springer Nature remains neutral with regard to jurisdictional claims in published maps and institutional affiliations.

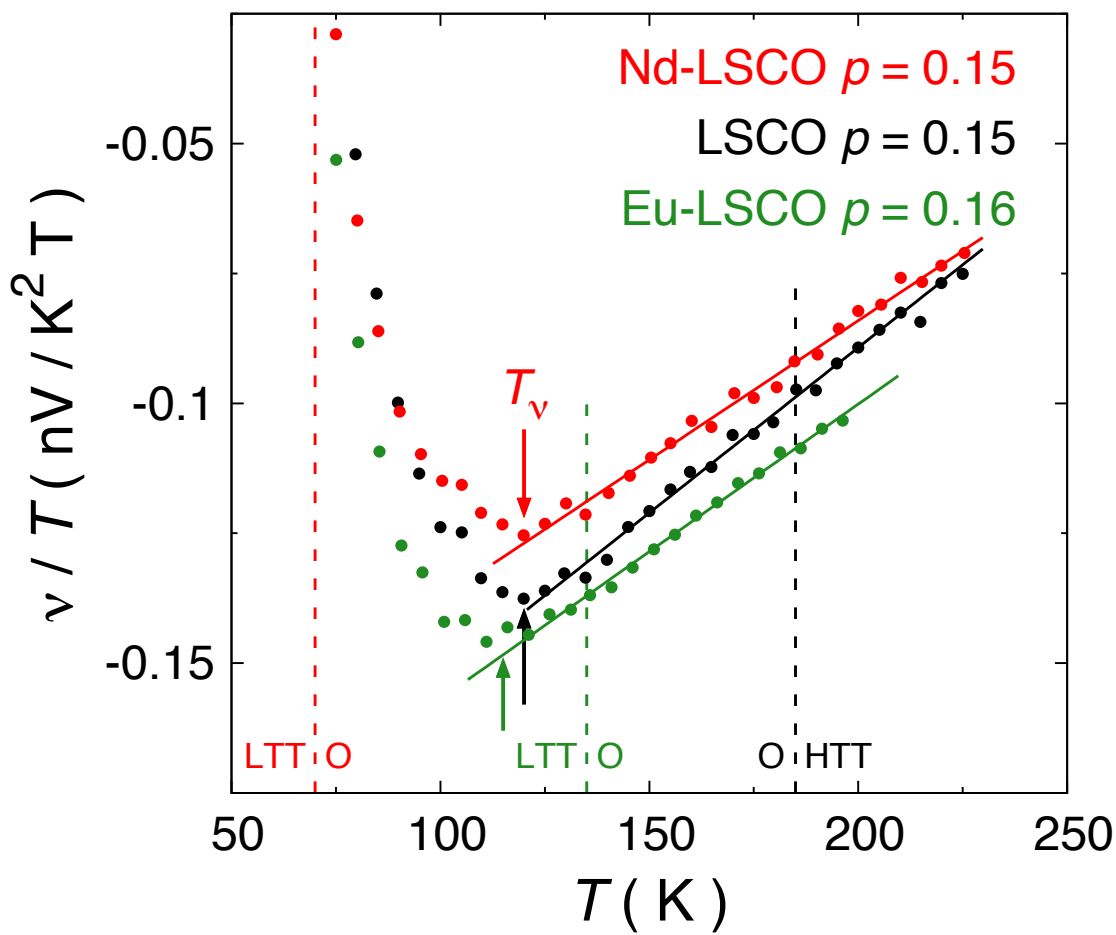

**Supplementary Figure 1 | Nernst coefficient for LSCO cuprates.**

Nernst coefficient $\nu$ vs temperature for LSCO $p = 0.15$ (black), Nd-LSCO $p = 0.15$ (red), and Eu-LSCO $p = 0.16$ (green), plotted as $\nu / T$ (figure adapted from ref. 1). The full lines are linear fits to the data and the arrows indicate the deviation from this linear behaviour at a temperature labeled $T_\nu$. The vertical dashed lines indicate the high temperature tetragonal (HTT) to orthorhombic (O) structural transition in LSCO (black), and the orthorhombic to low temperature tetragonal (LTT) transitions in Nd-LSCO (red) and Eu-LSCO (green). $T_\nu$ is essentially the same for all three materials, independent of where the structural transitions occur.

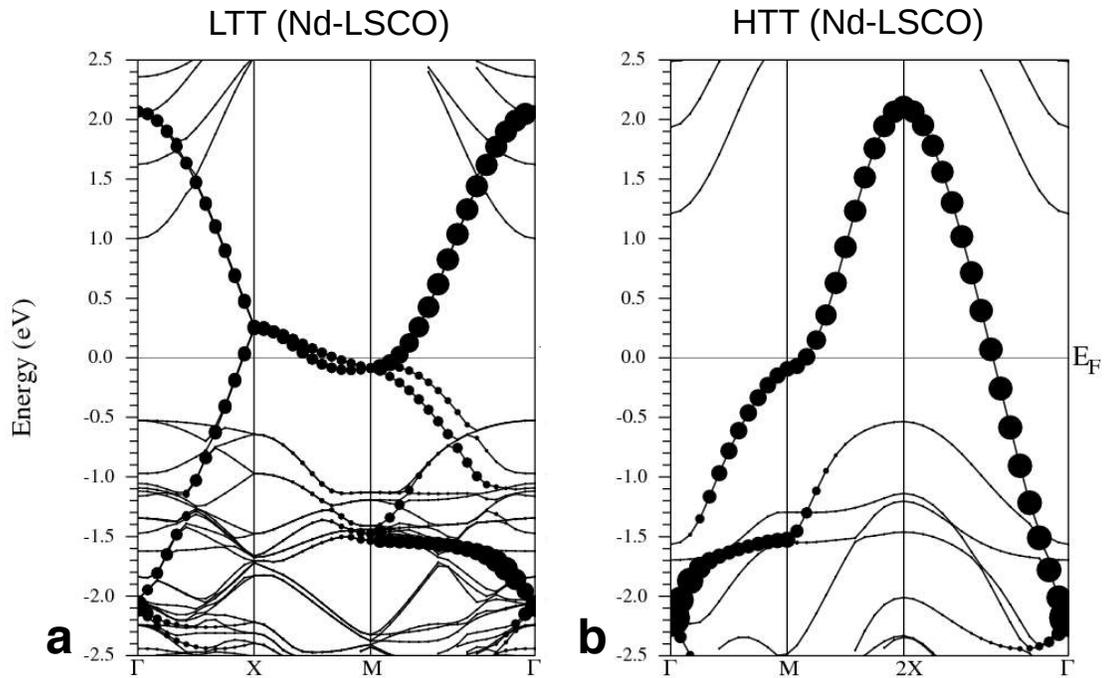

**Supplementary Figure 2 | Band structure calculations.**

Band-structure calculations for Nd-LSCO in the LTT structure at ambient pressure (**a**) and in the HTT structure at $P = 4.2$ GPa (**b**). The lattice parameters for the calculations are taken directly from X-ray diffraction data: ref. 2 for (a) and ref. 3 for (b). A tight-binding fit to the band structure gives the following values for the parameters $t$ and $t'$ : $t = 0.520$ eV, $t' = -0.089$, and $t' / t = -0.171$, for LTT ($P = 0$) ; $t = 0.539$ eV, $t' = -0.082$, and $t' / t = -0.152$, for HTT ($P = 4.2$ GPa).

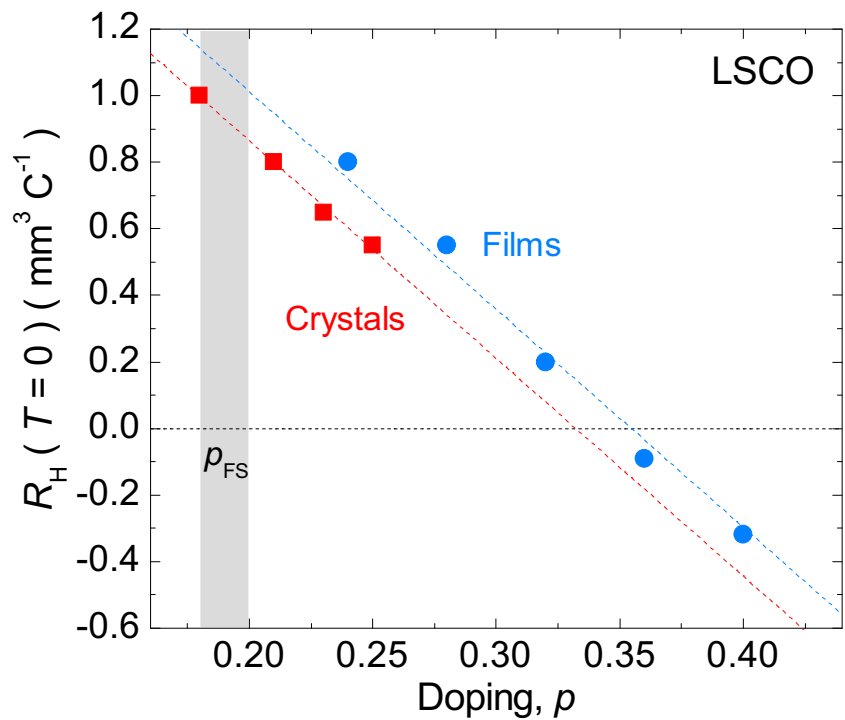

**Supplementary Figure 3 | Hall coefficient vs doping in LSCO.**

Hall coefficient $R_H$ of LSCO in the $T = 0$ limit, as a function of doping $p$, for $p = 0.18$ and higher. Data are based on measurements made on crystals (red squares from ref. 4) and thin films (blue dots, from ref. 5). Lines are linear fits through the respective data. $R_H$ is seen to decrease linearly as the doping increases away from $p_{FS}$ (grey band). We use this linearity of $R_H$ as a function of $p$ to quantify the shift of $p_{FS}$ with pressure in Nd-LSCO, in Figs. 4 and 5.

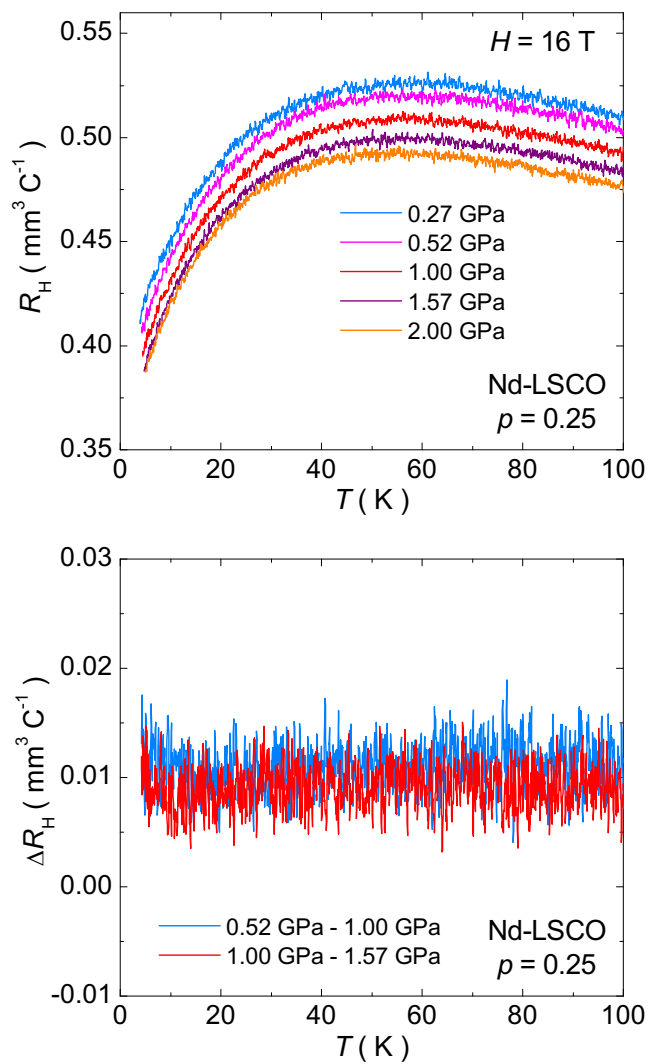

**Supplementary Figure 4 | Effect of pressure on the Hall coefficient of Nd-LSCO at $p > p^*$.**

**a)** Hall coefficient $R_H$ of Nd-LSCO at $p = 0.25$ (above $p^* = 0.23$) as a function of temperature, measured at $H = 16$ T, for various applied pressures as indicated.
**b)** Difference between two adjacent isobars in (**a**), for pairs as indicated. $R_H(T)$ decreases rigidly, at a rate of - 6 % $GPa^{-1}$, relative to $R_H(0) = 0.4$ mm$^3$ C$^{-1}$.

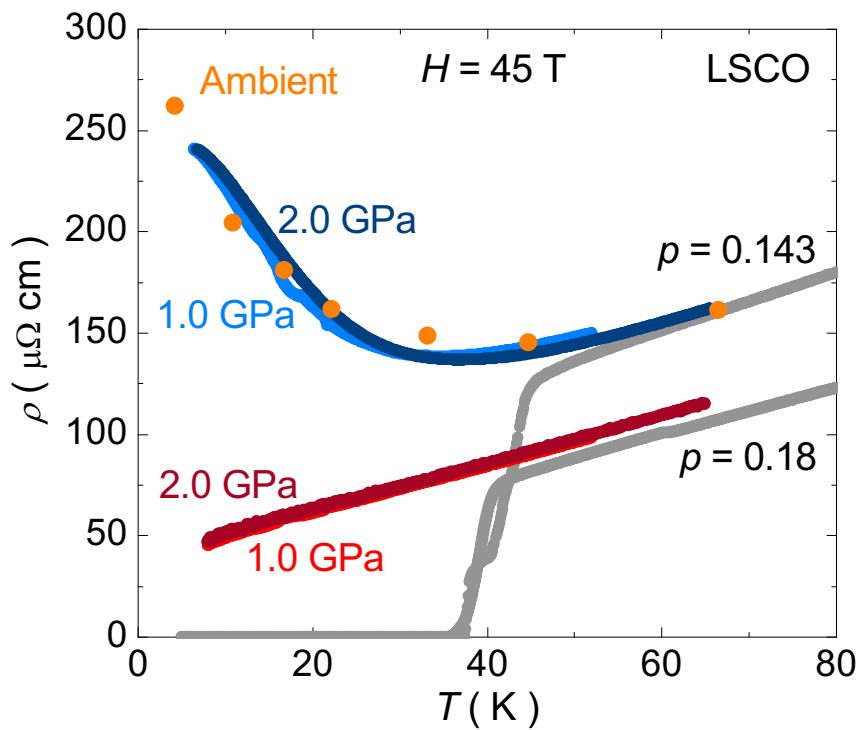

**Supplementary Figure 5 | Resistivity of LSCO under pressure.**

Resistivity of LSCO as a function of temperature, in $H$ = 45 T, for $p$ = 0.143 (blue and dark blue) and $p$ = 0.18 (red and burgundy), at pressures as indicated. Ambient pressure data for the same dopings, from ref. 6, are also shown, in grey for $H$ = 0 and orange for $H$ = 55 T.

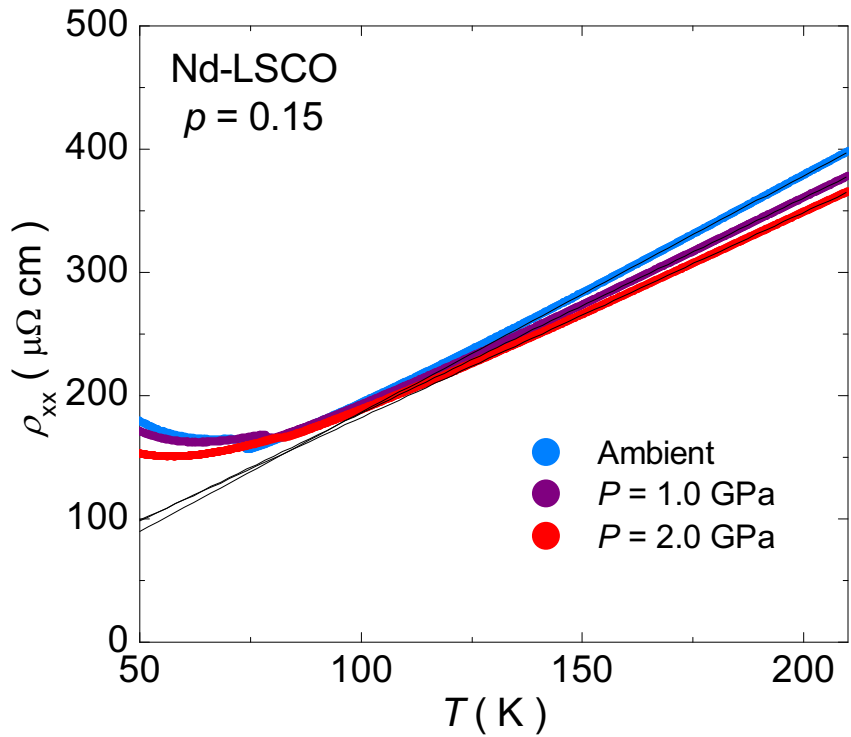

**Supplementary Figure 6 | Resistivity of Nd-LSCO at *p* = 0.15.**

Resistivity of Nd-LSCO as a function of temperature, at *p* = 0.15 and in zero magnetic field, for three different pressures as indicated. Lines are linear fits at high temperature. The difference between data and fit is plotted in Fig. 2b.

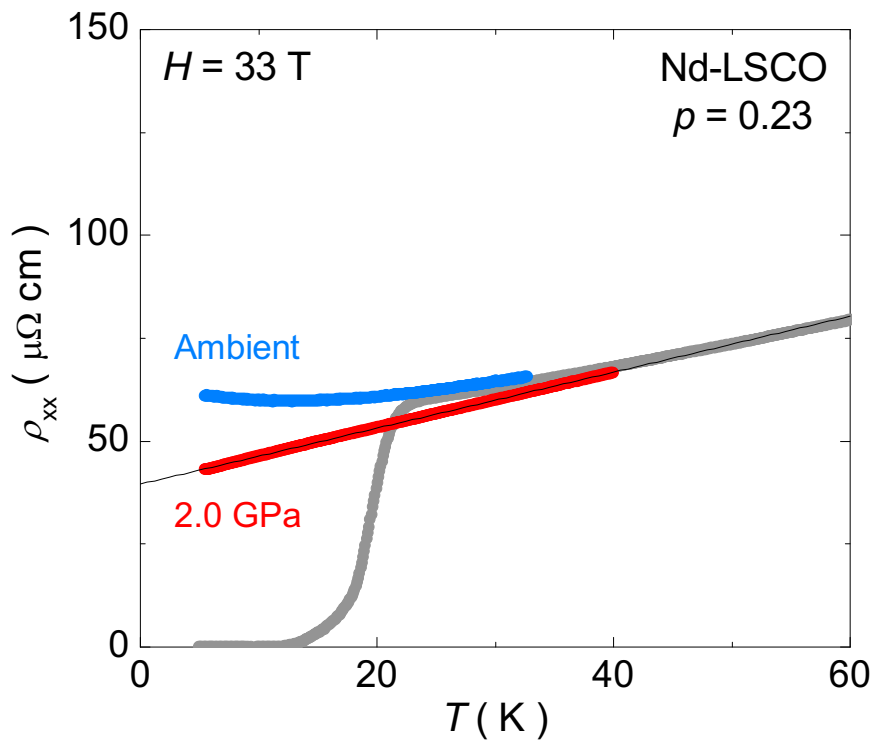

**Supplementary Figure 7 | Resistivity of Nd-LSCO at *p* = 0.23.**

Electrical resistivity of Nd-LSCO at *p* = 0.23 as a function of temperature, under ambient pressure (blue) and 2.0 GPa (red), in *H* = 33 T. Grey data are ambient pressure and zero field data on the same sample. The black line is a linear fit to the 2.0 GPa data.